\documentclass[preprint,onecolumn,nofootinbib]{revtex4}
\usepackage[colorlinks=true,linkcolor=blue,urlcolor=blue,filecolor=black,citecolor=red,pdfstartview=FitV,pdftitle={},pdfsubject={},pdfkeywords={},pdfpagemode=None,bookmarksopen=true]{hyperref}
\usepackage{autobreak}
\usepackage{graphicx}
\usepackage{amsmath}
\usepackage{amsfonts}
\usepackage{amssymb,ulem}
\usepackage{color}%
\usepackage{tikz}
\usepackage{dcolumn}
\usepackage{enumerate}
\usepackage{xcolor}

\allowdisplaybreaks
\begin{document}

\title{
  The mixed-state entanglement in holographic p-wave superconductor model
}
\author{Zhe Yang $^{1}$}
\email{yzar55@stu2021.jnu.edu.cn}
\author{Fang-Jing Cheng $^{2}$}
\email{fjcheng@mail.bnu.edu.cn}
\author{Chao Niu $^{1}$}
\email{niuchaophy@gmail.com}
\author{Cheng-Yong Zhang $^{1}$}
\email{zhangcy@email.jnu.edu.cn}
\author{Peng Liu $^{1}$}
\email{phylp@email.jnu.edu.cn}
\thanks{Corresponding author}

\affiliation{
  $^1$ Department of Physics and Siyuan Laboratory, Jinan University, Guangzhou 510632, China\\
  $^2$ Department of Astronomy, Beijing Normal University, Beijing 100875, China
}

\begin{abstract}

  In this paper, we investigate the mixed-state entanglement in a model of p-wave superconductivity phase transition using holographic methods. We calculate several entanglement measures, including holographic entanglement entropy (HEE), mutual information (MI), and entanglement wedge cross-section (EWCS). Our results show that these measures display critical behavior at the phase transition points, with the EWCS exhibiting opposite temperature behavior compared to the HEE. Additionally, we find that the critical exponents of all entanglement measures are twice those of the condensate. Moreover, we find that the EWCS is a more sensitive indicator of the critical behavior of phase transitions than the HEE. Furthermore, we uncover a universal inequality in the growth rates of EWCS and MI near critical points in thermal phase transitions, such as p-wave and s-wave superconductivity, suggesting that MI captures more information than EWCS when a phase transition first occurs.

\end{abstract}
\maketitle
\tableofcontents

\section{Introduction}\label{sec:intro}

Quantum entanglement is the most crucial characteristic of the quantum system and lays the key foundation of quantum information theory. Recently, quantum information has been attracting heavy attention from numerous fields, such as holographic theory, quantum many-body systems, and condensed matter theory. According to recent research, quantum information can detect quantum phase transitions and play a key role in spacetime emergence \cite{eisert2006entanglement,Osterloh:2002na,Amico:2007ag,Wen:2006topo,Kitaev:2006topo}.

In recent years, a variety of measures of quantum information have been proposed, such as entanglement entropy (EE), mutual information (MI), and R\'enyi entropy. EE is a widely used quantity that describes the entanglement of pure states very well. However, EE is not suitable for describing the entanglement of the more prevalent mixed states. To address this issue, new measures such as entanglement of purification (EOP), reflected entropy, quantum discord, and others have been suggested for mixed-state systems \cite{Vidal:2002zz,Horodecki:2009zz}. However, calculating these measures of quantum information can be challenging, particularly in strongly correlated systems. The complexity of these calculations increases exponentially with the size of the quantum system.

The gauge/gravity duality theory has been proved powerful tool for studying strongly correlated quantum systems by dualizing such systems to classical gravitational systems \cite{tHooft:1993dmi,Susskind:1994vu,Maldacena:1997re,Witten:1998qj,Hartnoll:2016apf}. It has been shown that the background geometry of the dual gravitational system encodes the quantum information of the dual field theory. For instance, the entanglement entropy (EE) is related to the minimum surface in the bulk, also known as the holographic entanglement entropy (HEE) \cite{Ryu:2006bv}. The ability of HEE to detect quantum phase transitions and thermal phase transitions has been investigated in \cite{Zeng:2015tfj,Cai:2012nm,Peng:2014ira,Peng:2015yaa}.
Recently, the entanglement wedge cross-section (EWCS) has been proposed as a novel measure of mixed-state entanglement in holographic systems \cite{Takayanagi:2017knl,Umemoto:2018jpc}. Additionally, various types of mixed-state entanglement, such as reflected entropy, logarithmic negativity, balanced partial entanglement, and odd entropy have been linked to the EWCS in holographic systems \cite{Kudler-Flam:2018qjo,Dutta:2019gen,Jokela:2019ebz,BabaeiVelni:2019pkw,Vasli:2022kfu,Camargo:2022mme}. In conclusion, EWCS is a powerful tool for investigating mixed-state entanglement in strongly coupled field theories \cite{Liu:2019qje,Huang:2019zph,Liu:2020blk,Chen:2021bjt,Li:2021rff,Chowdhury:2021idy,Sahraei:2021wqn,ChowdhuryRoy:2022dgo,Maulik:2022hty,Jain:2020rbb,Jain:2022hxl}.

Holographic superconductivity is a key topic in the gauge/gravity theory, providing a novel approach to studying high-temperature superconductors \cite{Hartnoll:2008kx,Horowitz:2010gk,Hartnoll:2008vx,Ling:2014laa,Rogatko:2015awa}. The symmetry of the Cooper pair wave function allows for the classification of superconductors as s-wave, p-wave, d-wave, etc. The main characteristics of the phase transition in superconductors are spontaneous symmetry breaking and the emergence of order parameters. For instance, an s-wave holographic superconductor is thought to be the spontaneous scalarization of the black hole, a p-wave holographic superconductor requires a charged vector field in the bulk as the vector order parameter, and a d-wave model was built by introducing a charged massive spin two field propagating in the bulk \cite{Arias:2012py,Cai:2015cya,Rogatko:2015nta}. Recent studies have shown that holographic quantum information can be used to detect the phase transition of s-wave superconductor \cite{Liu:2020blk,Peng:2014ira,Albash:2012pd,Jeong:2022zea}. However, research on the effects of mixed-state entanglement in p-wave superconductors is currently lacking. Therefore, it would be interesting to investigate the connection between the holographic p-wave superconducting phase transition and mixed-state entanglement.

In this paper, we aim to systematically study the role of mixed-state entanglement during the p-wave superconductivity phase transition. The paper is organized as follows: In Sec. \ref{sec1}, we introduce the holographic p-wave superconductor model, and the concepts of holographic quantum information, including holographic HEE, MI and EWCS. We explore the characteristics of mixed-state entanglement in Sec. \ref{sec2}. In Sec. \ref{sec3}, we provide analytical and numerical analysis of the scaling behavior of mixed-state entanglement measures. Additionally, we uncover an inequality between EWCS and MI in Sec. \ref{sec4}. Finally, in Sec. \ref{sec5}, we summarize our findings and conclusions.

\section{Holographic setup for p-wave superconductor and Holographic information-related quantities}\label{sec1}

We begin by presenting the model of a holographic p-wave superconductor and its phase diagram. Following that, we introduce HEE, as well as the mixed-state entanglement measures MI and EWCS.

\subsection{The holographic p-wave superconductor model}

In the p-wave superconductor model, as the temperature drops to a specific critical value, spontaneous symmetry breaking occurs, resulting in a vector order parameter. The system then transits from the normal phase (absence of vector hair) to the superconducting phase (presence of vector hair). The holographic p-wave model is constructed by introducing a complex vector field into Einstein-Maxwell theory with a negative cosmological constant \cite{Cai:2013aca,Cai:2014ija},
\begin{equation}\label{eq:action}
  \begin{aligned}
    S= & \frac{1}{2\kappa^2}\int d^4x\sqrt{-g}\left(\mathcal{R}+\frac{6}{L^2}-\frac{1}{4}F_{\mu\nu}F^{\mu\nu}-
    \frac{1}{2}\rho^\dagger_{\mu\nu}\rho^{\mu\nu}-m^2\rho_\mu^\dagger\rho^{\mu}+iq\gamma\rho_\mu\rho_\nu^\dagger F^{\mu\nu}\right),
  \end{aligned}
\end{equation}
where $\kappa^2=8\pi G$ is related to the gravitational constant, $L$ the AdS radius that we set as $1$.
$A$ is the gauge field and the field strength $F_{\mu\nu}=\nabla_{\mu}A_\nu - \nabla_\nu A_{\mu}$. $\rho_\mu$ is a complex vector field with mass $m$ and charge $q$. The tensor $\rho_{\mu\nu}=D_\mu\rho_\nu-D_\nu\rho_\mu$ with covariant derivative defined as $D_\mu=\nabla_\mu-iqA_\mu$. The last term in the action is the non-minimum coupling term between the Maxwell field and the complex vector field. In this paper, we only consider the case without an external magnetic field. The equation of motion (EOM) can be read as,
\begin{equation}\label{eq:eom}
  \begin{aligned}
    \nabla^\nu F_{\nu\mu}                                                          = & iq(\rho^\nu\rho^\dagger_{\nu\mu}-\rho^{\nu\dagger}\rho_{\nu\mu})+iq\gamma\nabla^\nu(\rho_\nu\rho_\mu^\dagger-\rho_\nu^\dagger\rho_\mu),                                                                                          \\
    D^\nu\rho_{\nu\mu}                                                             - & m^2\rho_\mu+iq\gamma\rho^\nu F_{\nu\mu}=0,                                                                                                                                                                                       \\
    \mathcal{R}_{\mu\nu}-\frac{1}{2}\mathcal{R}g_{\mu\nu}-\frac{3}{L^2}g_{\mu\nu} =  & \frac{1}{2}F_{\mu\lambda}F_\nu^\lambda+\frac{1}{2}\left(-\frac{1}{4}F_{\mu\nu}F^{\mu\nu}-\frac{1}{2}\rho_{\mu\nu}^\dagger\rho^{\mu\nu}-m^2\rho_\mu^\dagger\rho^\mu+iq\gamma\rho_\mu\rho_\nu^\dagger F^{\mu\nu}\right)g_{\mu\nu}+ \\
                                                                                     & \frac{1}{2}\{[\rho^\dagger_{\mu\lambda}\rho_\nu^\lambda+m^2\rho_\mu^\dagger\rho_\nu-iq\gamma(\rho_\mu\rho_\lambda^\dagger-\rho_\mu^\dagger\rho_\lambda)F_\nu^\lambda]+\mu\leftrightarrow\nu\}.
  \end{aligned}
\end{equation}
We solve the EOM with this ansatz,
\begin{equation}\label{eq:ansatz2}
  \begin{aligned}
    ds^2 & =\frac{1}{z^2}\left(-p(z)( 1-z)U(z)dt^2+\frac{1}{p(z)(1-z)U(z)}dz^2+V_1(z)dx^2+V_2(z)dy^2\right), \\
         & A_\nu dx^\nu=\mu(1-z)a(z)dt,\qquad \rho_\nu dx^\nu=\rho_x(z)dx,
  \end{aligned}
\end{equation}
where $p(z)\equiv 1+z+z^2-\frac{\mu^2z^3}{4}$. $\mu$ is the chemical potential of the dual field theory. The radius axis is denoted by $z$, which ranges from $0$ to $1$, with $z=0$ and $z=1$ representing the AdS boundary and horizon, respectively. In our ansatz, there are five unknown functions, $U(z)$, ${V_1}(z)$, ${V_2}(z)$, $a(z)$, and $\rho_x(z)$, which can be obtained by solving the EOM. The ansatz \eqref{eq:ansatz2} reduces to the AdS-RN black brane solution when $U=V_1=V_2=a=1$ and $\rho_x=0$.
The expansion of the $\rho_x$ near the AdS boundary is
\begin{equation}
  \rho_x=\rho_{x_-}z^{\Delta_-}+\rho_{x_+}z^{\Delta_+}+\cdots,
\end{equation}
where the scaling dimension $\Delta_{\pm}=\frac{1\pm\sqrt{1+4m^2}}{2}$ and we set the source $\rho_{x_-}=0$ for the condensate arise spontaneously. After solving the EOM, we can obtain the condensate $\langle J_x\rangle$ by extracting the coefficient $\rho_{x_+}$ . The condensate $\langle J_x\rangle$ emerges at a specific temperature when varying $m^2$ and $q$. Consequently, in the dual quantum field, the vector operator acquires a non-zero vacuum expectation value and spontaneously breaking the U(1) symmetry and rotational symmetry. Therefore, $\langle J_x\rangle$ can be used as the order parameter of p-wave superconducting phase transition.

The Hawking temperature of this model is $\tilde{T}=\frac{12-\mu^2}{16\pi}$. The system is invariant under the following rescaling,
\begin{equation}
  \begin{aligned}
    (t,x,y)\to\alpha^{-1}(t,x,y) & ,\quad (U,V_1,V_2)\to\alpha^2(U,V_1,V_2),                                            \\
    \mu\to\alpha\mu              & ,\quad \tilde{T}\to\alpha\tilde{T},\quad \rho_{x_+}\to\alpha^{\Delta_++1}\rho_{x_+}.
  \end{aligned}
\end{equation}
In this paper, we adopt the chemical potential $\mu$ as the scaling unit, which is equivalent to treating the dual system as a field theory described by the giant canonical ensemble. The dimensionless Hawking temperature $T=\tilde{T}/\mu$.

\subsection{The phase diagram of holographic p-wave superconductor model}

\begin{figure}
  \centering
  \includegraphics[width=0.45\textwidth]{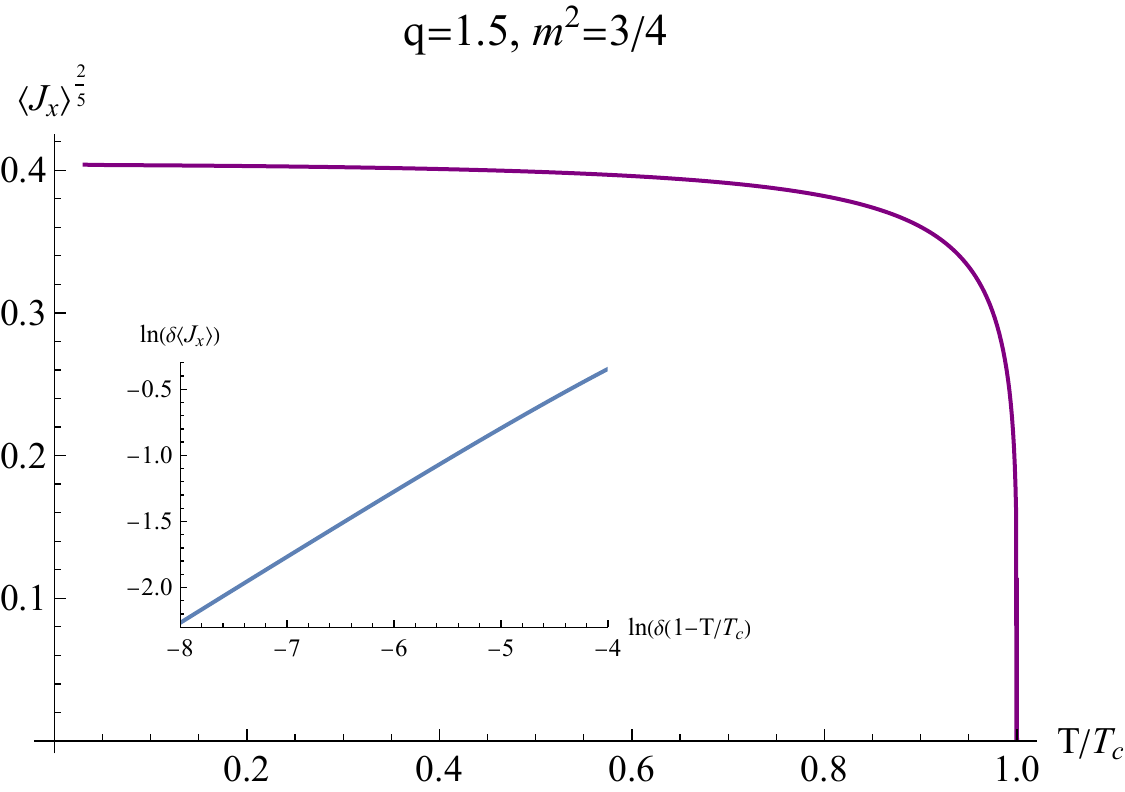}\qquad
  \includegraphics[width=0.45\textwidth]{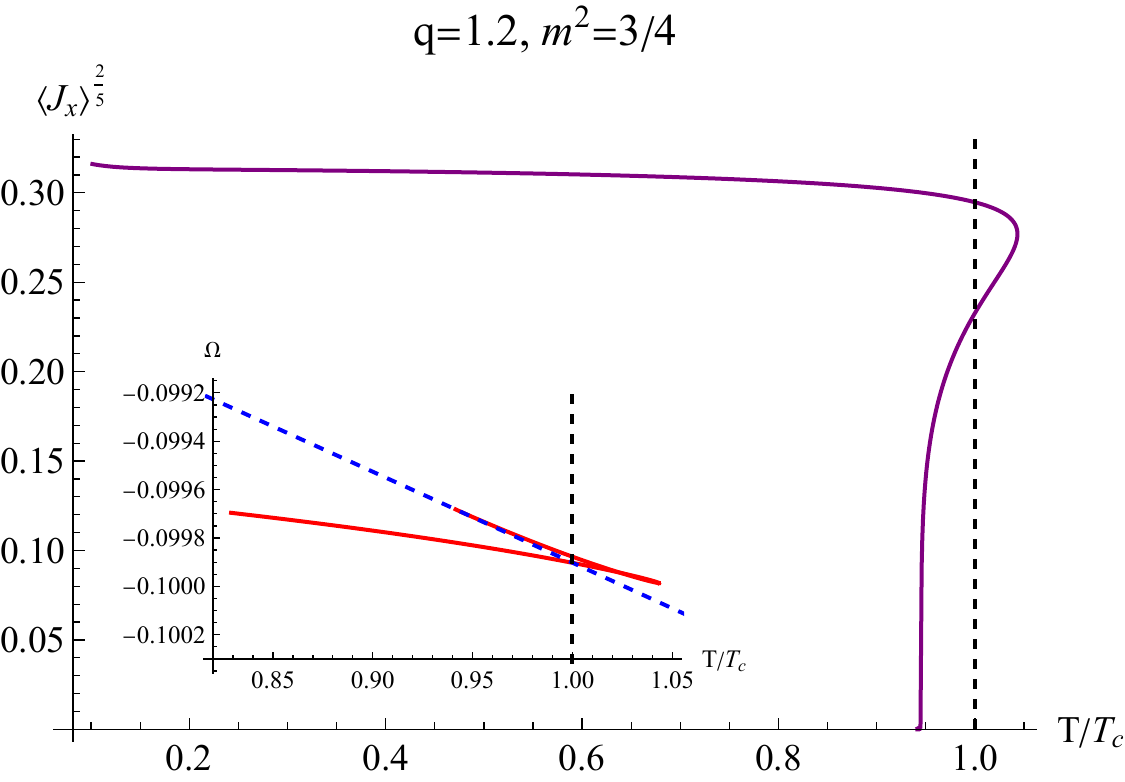}
  \caption{Left plot: The second-order phase transition occurs as the temperature falls below the critical value. The inset plot illustrates the scaling behavior of the condensate $\langle J_x\rangle$. Right plot: The first-order phase transition occurs when the temperature falls below the critical temperature, which represents by the black dashed line. The inset plot illustrates the effective free energy density $\Omega$ versus the temperature $T/T_c$, revealing that the superconducting phase is thermodynamically favored.}
  \label{fig1}
\end{figure}

This holographic p-wave superconductor model can exhibit zeroth-order, first-order, and second-order phase transitions depending on the values of $m$ and $q$. For example, a second-order phase transition can occur at $q=1.5$, $m^2=3/4$ with a critical temperature of $T_c\approx 0.01791$. In Fig. \ref{fig1}, we demonstrate the relationship between the condensate $\langle J_x\rangle^{2/5}$ and temperature by plotting the scaling relationship,
\begin{equation}
  \delta(\langle J_x \rangle)\sim\left(1-\frac{T}{T_c}\right)^{\alpha_{c}}.
\end{equation}
Theoretical calculations predict that the critical exponent is $\alpha=1/2$ \cite{Cai:2013aca}. Our numerical results also indicate that $\alpha_c\approx0.500106$.

A first-order phase transition can occur at $q=1.2$ and $m^2=3/4$ with a critical temperature of $T_c\approx 0.003382$. To better visualize the phase structure, we plot the effective free energy density (as shown in Fig.\ref{fig1}). The effective free energy density is defined as $\tilde{\Omega}=M-Ts$, where $T$ is the Hawking temperature, $s$ is the entropy density, and $M$ is the mass density of the black brane \cite{Liu:2021rks}. The mass density of the black brane can be obtained by using the AdS asymptotic behavior of $g_{tt}$ in our ansatz,
\begin{equation}
  \frac{(1-z) U(z) \left(-\frac{1}{4} \mu ^2 z^3+z^2+z+1\right)}{z^2}\sim \frac{1}{z^2}+M z+ Q z^2+\cdots.
\end{equation}
The free energy of the superconducting phase is lower than the normal phase when the temperature drops below the critical temperature $T_c$. As a result, the system will abruptly transition from the normal phase to the superconducting phase.

\begin{figure}
  \centering
  \includegraphics[width=0.6\textwidth]{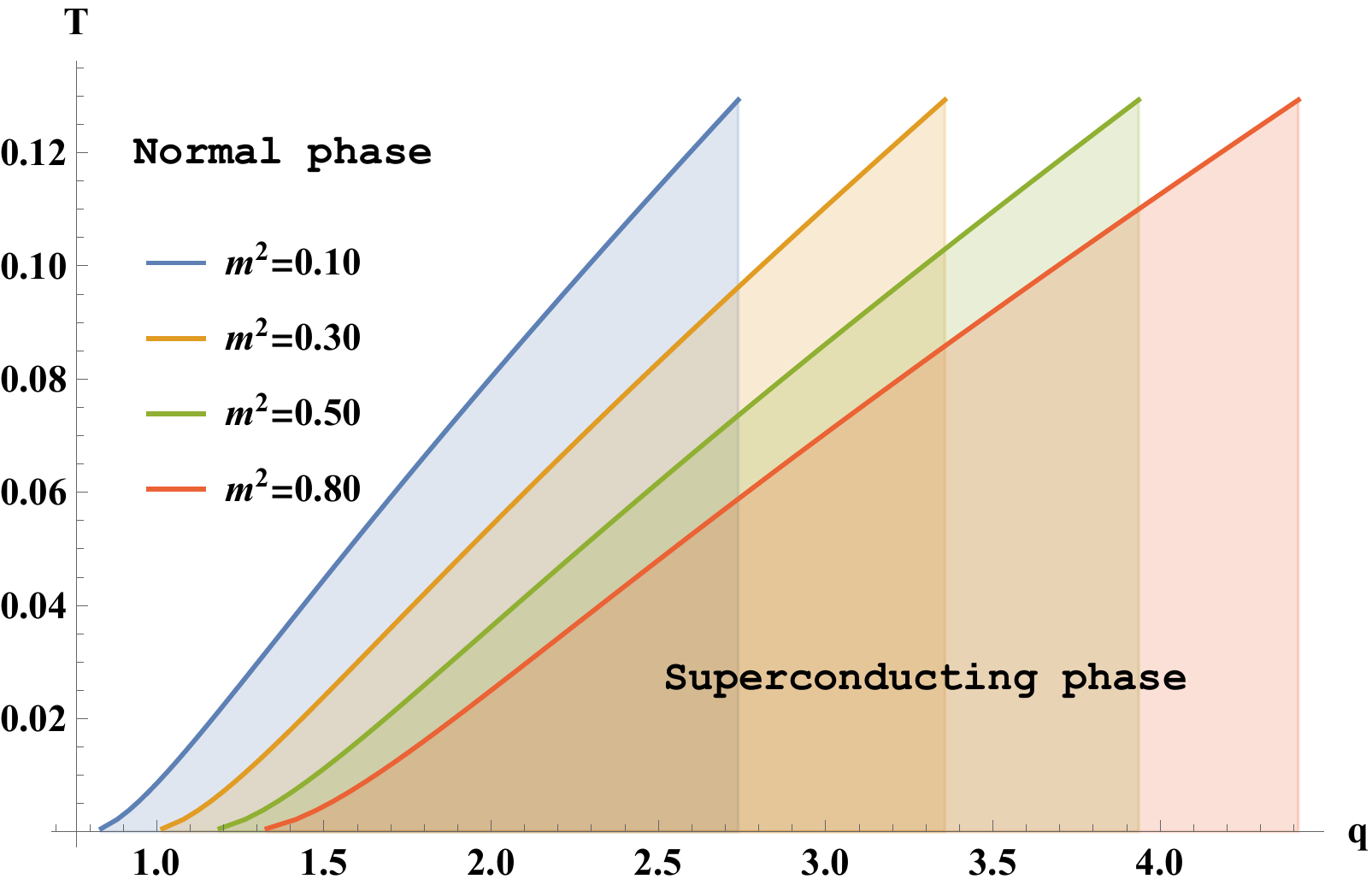}
  \caption{The phase diagram of holographic p-wave superconductor model with positive $m^2$. The solid lines are the critical points.}
  \label{fig14}
\end{figure}

To more thoroughly understand the behavior of p-wave superconductivity, we present the phase diagram in Fig. \ref{fig14}. The phase diagram is constructed by identifying the critical points, which can be found by examining the emergence of condensation as a perturbation near these points. The linearized equations of motion can be transformed into an eigenvalue problem that we solve using numerical methods
\begin{equation}\label{eq:phasediagram}
  \begin{aligned}
    \frac{1}{32 \mu ^2 (z-1) z^2}(\mu ^2 z^3-4 z^2-4 z  & -4) (2 z^2 (z^2 (\mu ^2 (4 z-3)-12) \rho_x '(z)+ \\
    \left(\mu ^2 z^4-\left(\mu ^2+4\right) z^3+4\right) & \rho_x ''(z))-6 \rho_x (z))=-q^2 \rho_x (z).
  \end{aligned}
\end{equation}
By analyzing the eigenvalues, we can determine the upper or lower bounds of the critical points, which correspond to the boundaries of the different phases in the diagram.

\subsection{The holographic quantum information}

Quantum entanglement is a fundamental characteristic of quantum systems. EE is a well-known measure of entanglement, which quantifies the correlation between a subsystem and its complement for pure states. It is defined in terms of the reduced density matrix $\rho_A$ \cite{Eisert:2008ur},
\begin{equation}
  S_A(|\psi\rangle)=-\text{Tr}[\rho_A\text{log}(\rho_A)],\qquad \rho_A=\text{Tr}_B(|\psi\rangle\langle\psi|).
\end{equation}

\begin{figure}
  \centering
  \begin{tikzpicture}[scale=1]
    \node [above right] at (0,0) {\includegraphics[width=8cm]{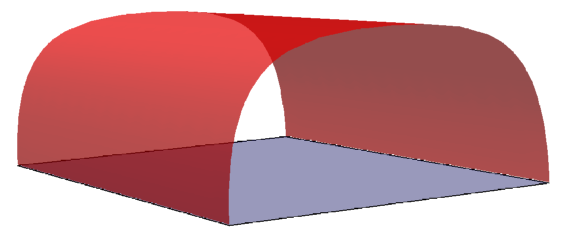}};
    \draw [right,->,thick] (3.4, 0.2) -- (5.25, 0.5) node[below] {$x$};
    \draw [right,->,thick] (3.4, 0.2) -- (1.85, 0.7) node[below] {$y$};
    \draw [right,->,thick] (3.4, 0.2) -- (3.4, 3.425) node[above] {$z$};
  \end{tikzpicture}
  \begin{tikzpicture}[scale=1]
    \node [above right] at (0,0) {\includegraphics[width=7.3cm]{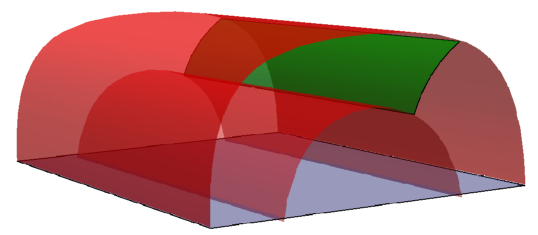}};
    \draw [right,->,thick] (2.94, 0.221) -- (4.55, 0.45) node[below] {$x$};
    \draw [right,->,thick] (2.94, 0.221) -- (1.45, 0.75) node[below] {$y$};
    \draw [right,->,thick] (2.94, 0.221) -- (3.0, 3.425) node[above] {$z$};
  \end{tikzpicture}
  \caption{Left plot: the minimum surface for a subsystem (red region). Right plot: the minimum cross-section (green surface) of the entanglement wedge.}
  \label{fig3}
\end{figure}
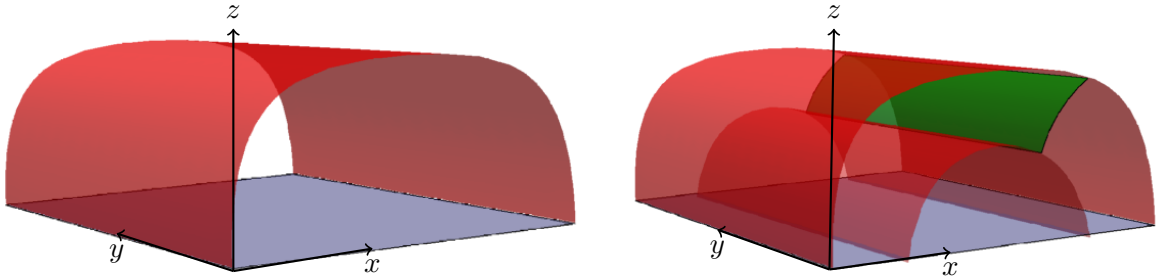

The HEE was proposed to be dual to the area of the minimum surface in the gravitational system \cite{Nishioka:2009un}.
In this paper, we consider the HEE of the configuration with an infinitely long strip along the $y$-axis (see Fig. \ref{fig3}). HEE typically diverges due to the asymptotic AdS boundary. The regulation is implemented by subtracting the divergent term from the HEE. It should be noted that HEE is not suitable for describing the mixed-state entanglement. For example, EE of the quantum system characterized by the direct product state $\mathcal{H}_A\otimes\mathcal{H}_B$ is not equal to zero, but the entanglement of the subsystems is vanishing. This is because EE contains both quantum and classical correlation. Therefore, as the dual of EE, HEE is also affected by thermodynamic entropy in mixed-state systems \cite{Ling:2015dma,Ling:2016wyr}.

\begin{figure}
  \centering
  \includegraphics[width=0.5\textwidth]{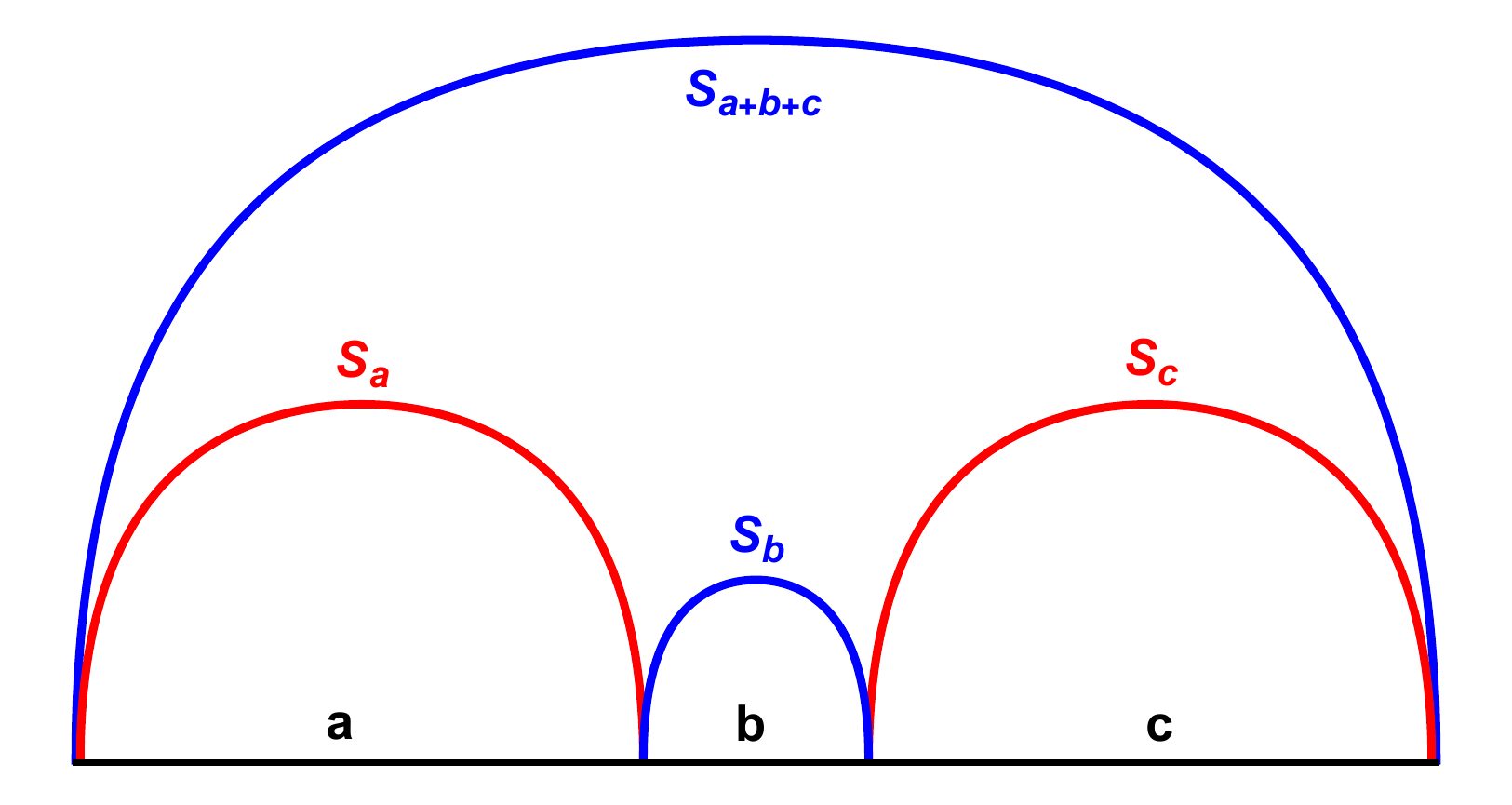}
  \caption{The illustration of the holographic mutual information.}
  \label{fig4}
\end{figure}

To better solve the problem of mixed-state entanglement measurement, numerous novel entanglement measures have been proposed.
One popular measure is mutual information (MI), which quantifies the correlation between two subsystems $A$ and $C$ that are separated by a subsystem $B$. According to the definition of MI, it is calculated as \cite{chuang:2002,Hayden:2011ag},
\begin{equation}\label{eq:midefine}
  I(a:c)=S(a)+S(c)-\text{min}(S(a\cup c)),
\end{equation}
where $S(x)$ denotes the entanglement entropy of subsystem $x$. Unlike entanglement entropy, MI for direct product states $\mathcal{H}_A\otimes\mathcal{H}_B$ is always zero, making it a more appropriate measure for mixed-state entanglement. In the holographic context, the dual of MI is the difference in area between red (disconnected configuration) and blue surfaces (connected configuration), as shown in Fig. \ref{fig4}. As the subsystem $A,\,C$ becomes smaller or when the separation $B$ becomes larger, MI decreases and eventually reaches zero, indicating a disentangling phase transition. However, MI has some limitations as a mixed-state entanglement measure as it is directly related to entanglement entropy and can be dominated by it in some cases \cite{Huang:2019zph}. Therefore, it is important to explore other mixed-state entanglement measures.

Recently, the minimum cross-section of the entanglement wedge (EWCS) is proposed as a novel holographic mixed-state entanglement measure \cite{Takayanagi:2017knl}. EWCS is considered to be the duality of reflected entropy, logarithmic negativity, and odd entropy. The definition of EWCS is as follows,
\begin{equation}
  E_w(\rho_{AB})=\underset{\Sigma_{AB}}{\text{min}}\left( \frac{\text{Area}(\Sigma_{AB})}{4G_N}  \right).
\end{equation}
Fig. \ref{fig3} is an illustration of EWCS in a bipartite system $a\cup c$ divided by $b$. The area bounded by the minimum surface of the disconnected configuration is known as the entanglement wedge. It is important to note that entanglement between subsystems only exists when the total correlation is not zero, which means the MI does not vanish.

\begin{figure}
  \centering
  \includegraphics[width=0.8\textwidth]{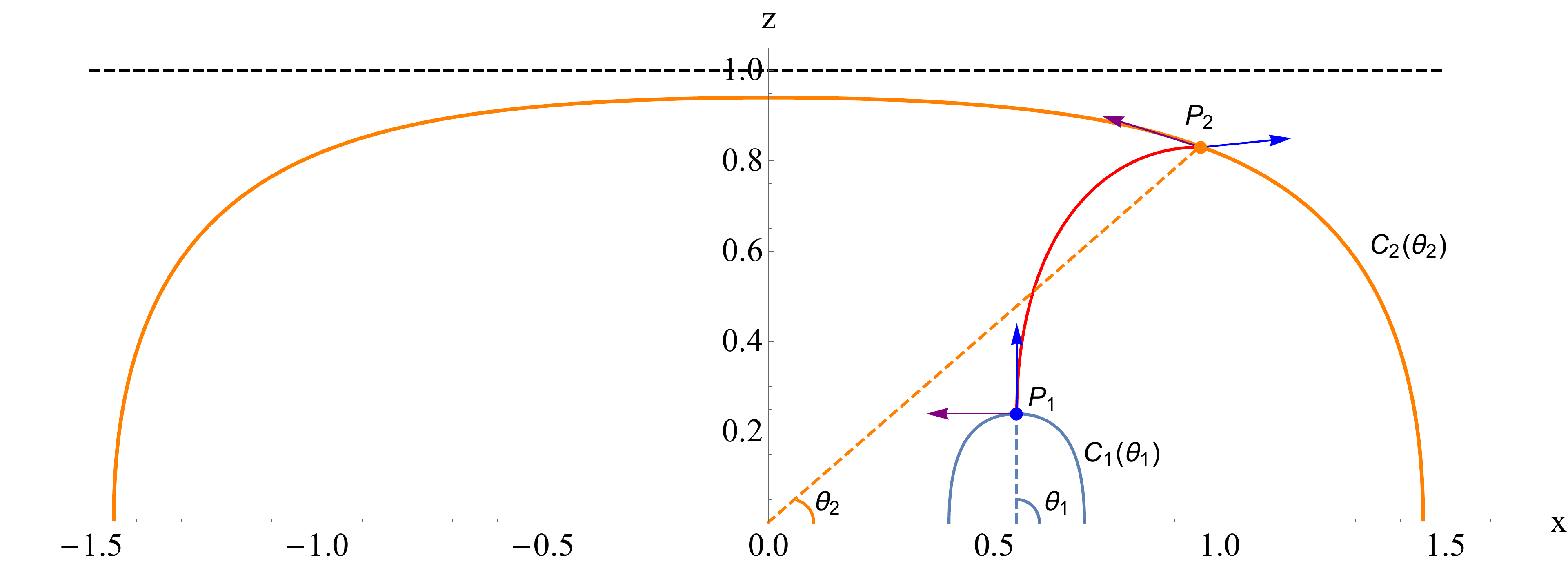}
  \caption{The illustration of the numerical algorithm for EWCS.}
  \label{fig5}
\end{figure}

Although EWCS plays a significant role in measuring the entanglement of mixed-state systems, it is still challenging to solve it \cite{Liu:2019qje}. First, it is hard to solve the highly nonlinear EOM of the minimum surface. Second, the minimum cross-section is obtained by scanning a two-dimensional parameter space, which is a hard task. Last but not least, the coordinate singularity close to the AdS boundary with the asymptotic AdS can hinder numerical precision. We have proposed an efficient algorithm for solving EWCS based on the requirement that the minimum cross-section is locally orthogonal to the boundaries of the entanglement wedge \cite{Liu:2020blk}. Fig. \ref{fig5} shows the illustration of the key concept for our numerical algorithm. We consider EWCS of the infinite strip along the $ y $-direction in a homogeneous background
\begin{equation}
  ds^2=g_{tt}dt^2+g_{zz}dz^2+g_{xx}dx^2+g_{yy}dy^2.
\end{equation}
The minimum surfaces of the connected configuration can be represented as $C_1(\theta_1)$ and $C_2(\theta_2)$. The minimum surfaces intersect with the cross-section at points $p_1$ and $p_2$, and the area of this local minimum surface (the red curve in Fig. \ref{fig5}) is,
\begin{equation}\label{eq:areamc}
  A=\int_{C_{p_1,p_2}}\sqrt{g_{xx}g_{yy}x'(z)^2+g_{zz}g_{yy}}dz.
\end{equation}
Variating \eqref{eq:areamc}, we obtain the EOM determining the local minimum surface,
\begin{equation}
  x'(z)^3\left (\frac{g_{xx}g_{yy}'}{2g_{yy}g_{zz}}+\frac{g_{xx}'}{2g_{zz}}\right )+x'(z)\left (\frac{g_{xx}'}{g_{xx}}+\frac{g_{yy}'}{2g_{yy}}-\frac{g_{zz}'}{2g_{zz}}\right )+x''(z)=0.
\end{equation}
Remind that the global minimum cross-section is locally orthogonal to the entanglement wedge, which implies that
\begin{equation}
  \left \langle \frac{\partial}{\partial_z},\frac{\partial}{\partial \theta_1}\right \rangle_{p_1}=0, \quad
  \left \langle \frac{\partial}{\partial_z},\frac{\partial}{\partial \theta_2}\right \rangle_{p_2}=0
\end{equation}
where $\langle \cdot,\cdot\rangle$ represents the vector product with metric $g_{\mu\nu}$. We can normalize the orthogonal relation,
\begin{equation}
  \label{eq1}
  Q_1(\theta_1,\theta_2) \equiv\left.\frac{\langle \frac{\partial}{\partial z},\frac{\partial}{\partial \theta_1}\rangle}{\sqrt{\langle \frac{\partial}{\partial z},\frac{\partial}{\partial z}\rangle \langle\frac{\partial}{\partial \theta_1},\frac{\partial}{\partial \theta_1}\rangle}} \right |_{p_1}=0,\quad
  Q_2(\theta_1,\theta_2) \equiv\left.\frac{\langle \frac{\partial}{\partial z},\frac{\partial}{\partial \theta_1}\rangle}{\sqrt{\langle \frac{\partial}{\partial z},\frac{\partial}{\partial z}\rangle \langle\frac{\partial}{\partial \theta_2},\frac{\partial}{\partial \theta_2}\rangle}} \right |_{p_2} =0.
\end{equation}
Finding the cross-section located at the minimum surface at $(\theta_1,\theta_2)$ where \eqref{eq1} is satisfied, we obtain the minimum cross-section. To this end, we adopt the Newton-Raphson method to locate the endpoints satisfying the local perpendicular conditions. Based on the above techniques, we can study the relationship between the holographic p-wave superconductor and the EWCS \cite{Liu:2020blk}.

\section{The computation of the holographic quantum information}\label{sec2}

\subsection{The holographic entanglement entropy and mutual information}

\begin{figure}
  \centering
  \includegraphics[width=0.45\textwidth]{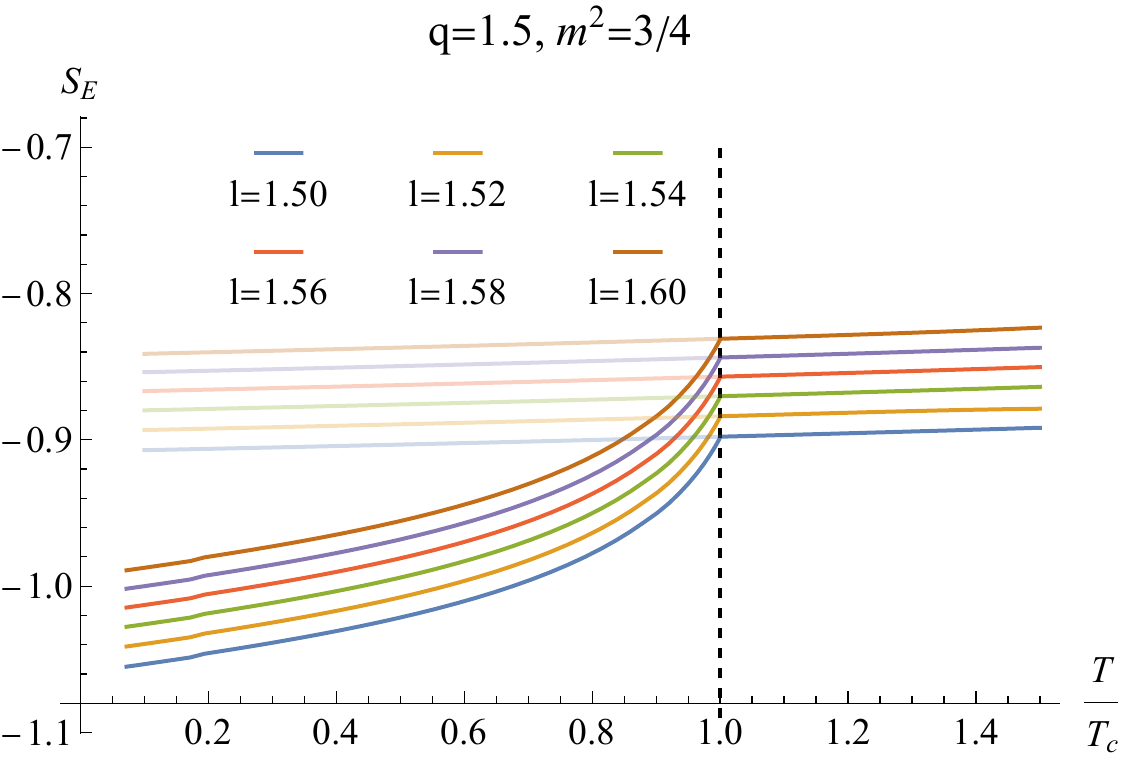}\qquad
  \includegraphics[width=0.45\textwidth]{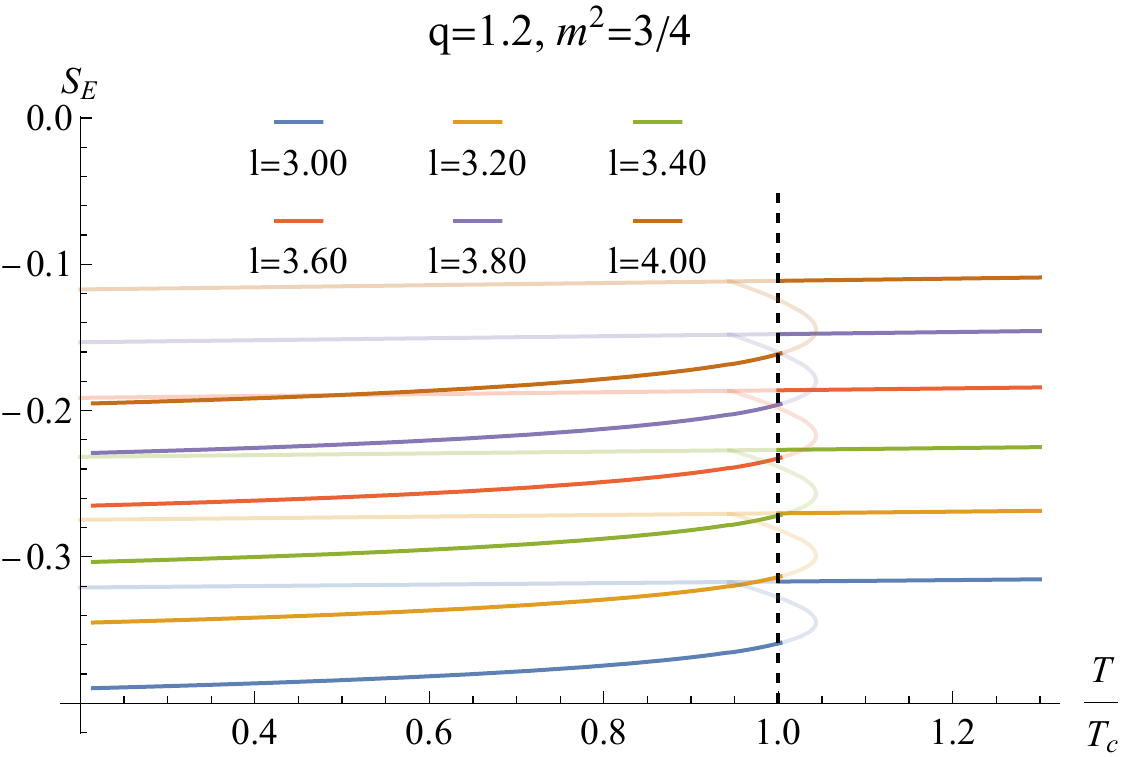}
  \caption{The holographic entanglement entropy $S_E$ vs temperature $T/T_c$ with various strip width $l$. The critical point is indicated by the black dashed line. The stable and metastable states are depicted by solid and transparent lines, respectively. Left plot: The second-order phase transition at $T_c\approx0.01791$. Right plot: The first-order phase transition at $T_c\approx0.003382$. }
  \label{fig6}
\end{figure}

In Fig. \ref{fig6}, we show the relationship between the HEE and temperature $T/T_c$ during second-order and first-order phase transitions. For $q=1.5$ and $m^2=3/4$, where the second-order phase transition occurs, the HEE increases with increasing temperature. For $q=1.2$ and $m^2=3/4$, where the first-order phase transition occurs, the HEE jumps abruptly when crossing the critical point. To understand this behavior, we can examine the relationship between HEE and thermodynamic entropy, as when the configuration is large or the temperature is high enough, the minimum surface will approach the horizon of the black brane and HEE will primarily be determined by thermodynamic entropy. Therefore, we will next analyze the thermodynamic entropy behavior of the black brane to better understand the behavior of HEE \cite{Ling:2016wyr,Ling:2015dma}.

\begin{figure}
  \centering
  \includegraphics[width=0.45\textwidth]{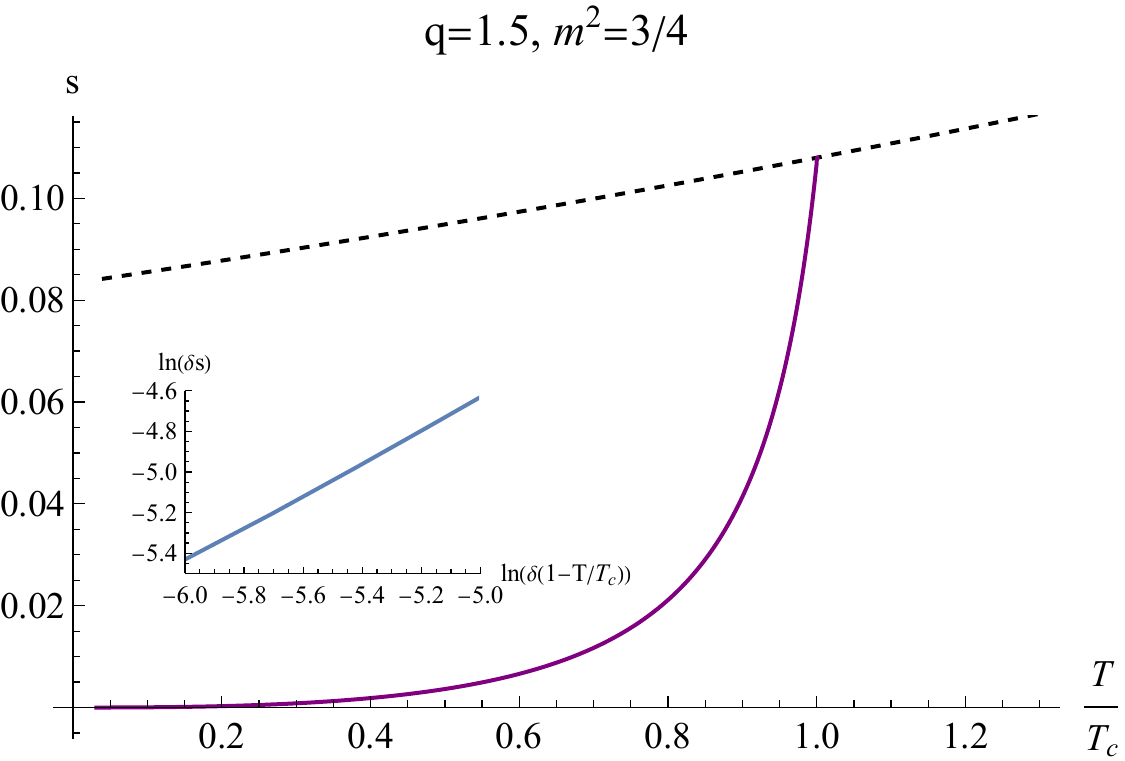}\qquad
  \includegraphics[width=0.45\textwidth]{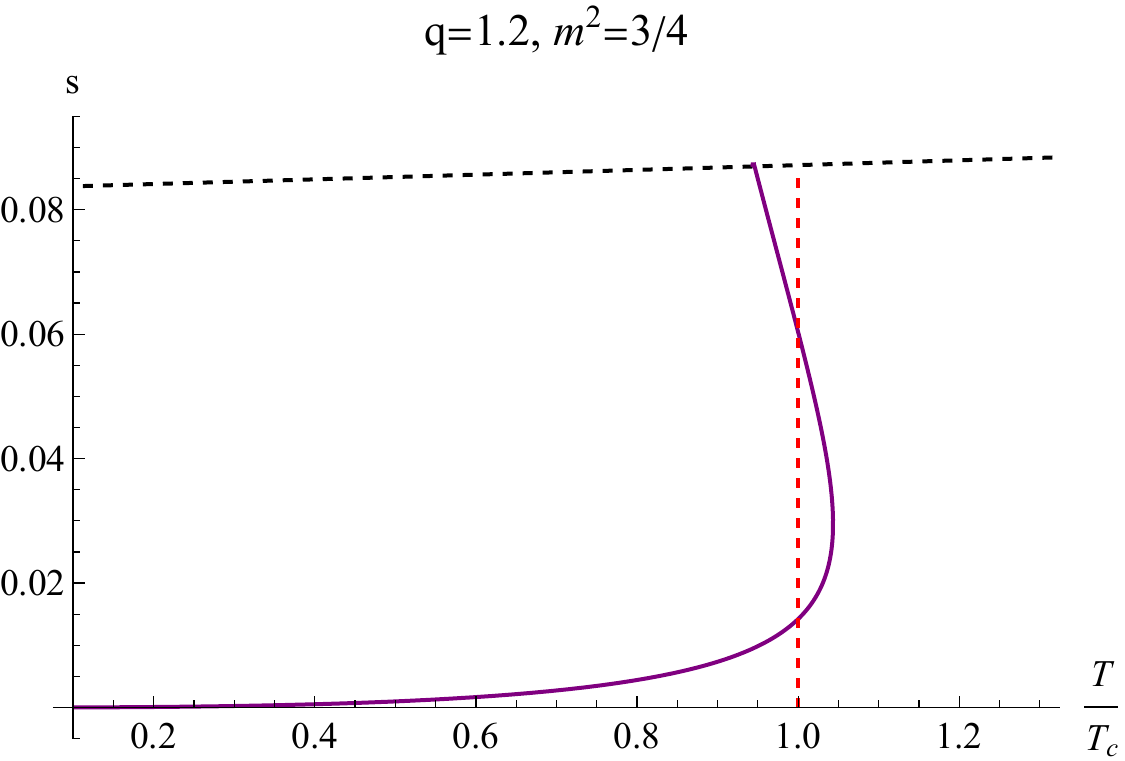}
  \caption{The entropy density $s$ vs temperature $T/T_c$. The black dashed line represents the entropy density of the normal phase. The entropy density of the superconducting phase is represented by the purple line. Left plot: The second-order phase transition of holographic p-wave superconductor model. The inset plot depicts the logarithm between $s$ and $1-\frac{T}{T_c}$. Right plot: The first-order phase transition of holographic p-wave superconductor model. The critical temperature is indicated by the red dashed line.}
  \label{fig7}
\end{figure}

The entropy density is given by,
\begin{equation}
  \tilde{s}=\frac{2\pi A}{\kappa^2}=\frac{2\pi \sqrt{V_1(z) V_2(z)} }{\kappa^2}\hat{V},
  \label{eqs}
\end{equation}
where $A$ is the area of the horizon and $\hat{V}=\int dxdy$ is the corresponding area of the region in the dual field theory \cite{Nguyen:2015wfa}. Dividing the entropy by the area $\hat{V}$ and $\mu^2$, we have the dimensionless entropy density $s=\frac{\kappa^2 \tilde{s}}{2\pi \hat{V} \mu^2}$. The plot of the entropy density near the critical point can be seen in Fig. \ref{fig7}. The above phenomena show that both HEE and entropy density can detect the critical behavior of the holographic p-wave superconducting phase transitions. Similar phenomena of the HEE in the superconducting phase transition can see in \cite{Liu:2020blk,Cai:2012sk,Peng:2015yaa,Peng:2014ira}.

\begin{figure}
  \centering
  \includegraphics[width=0.45\textwidth]{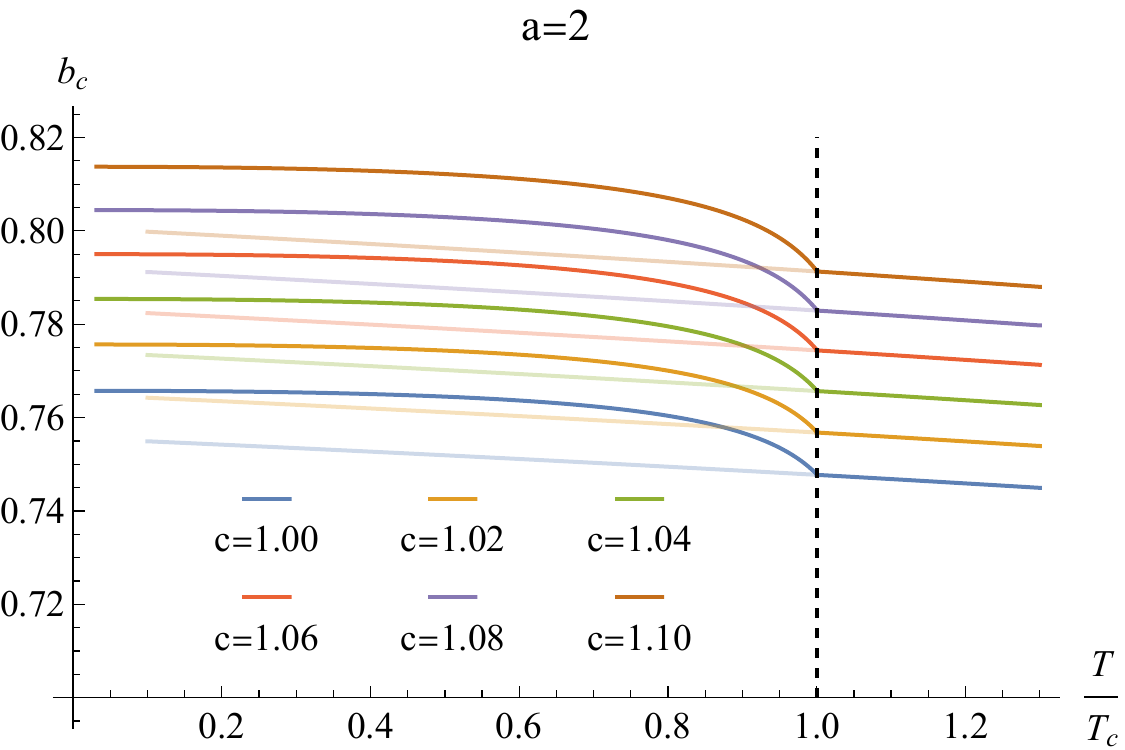}\qquad
  \includegraphics[width=0.45\textwidth]{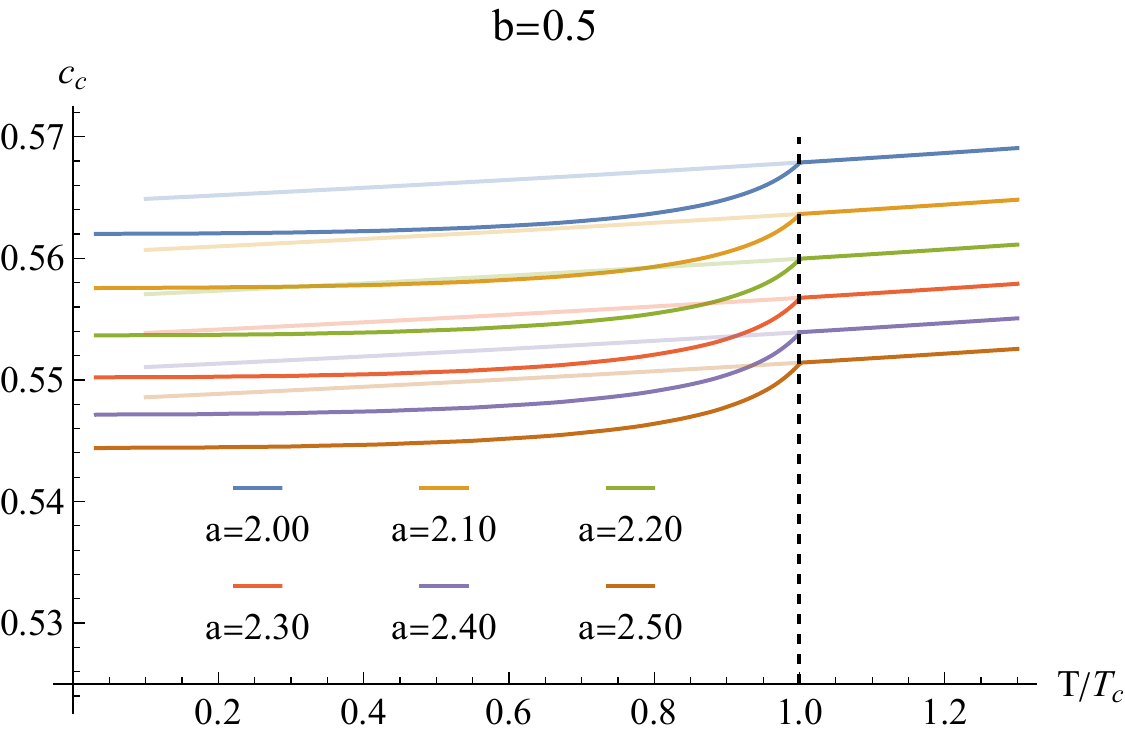}
  \caption{The critical configuration for disentangling phase transition. The critical temperature is indicated by the black dashed line. The solid and translucent lines represent stable and metastable states. Left plot: When $b$ is above the $b_c$, the disentangling phase transition occurs. Right plot: When $c$ is below $c_c$, the disentangling phase transition occurs.}
  \label{fig9}
\end{figure}

MI is one of the mixed-state entanglement measures that can extract the total correlation of the systems. Since MI is directly defined by HEE (see \eqref{eq:midefine}), it also can diagnose the phase transition. Moreover, a disentangling phase transition occurs when MI is zero and entanglement exists only when MI is greater than zero. Fig. \ref{fig9} illustrates the behavior of the disentangling phase transition for various configurations.
However, in certain cases, MI is determined by the thermodynamic entropy \cite{Liu:2020blk,Liu:2021rks,Huang:2019zph}. Therefore, it is necessary to investigate other mixed-state entanglement measures.

\subsection{The minimum entanglement wedge cross section}

\begin{figure}
  \centering
  \includegraphics[width=0.45\textwidth]{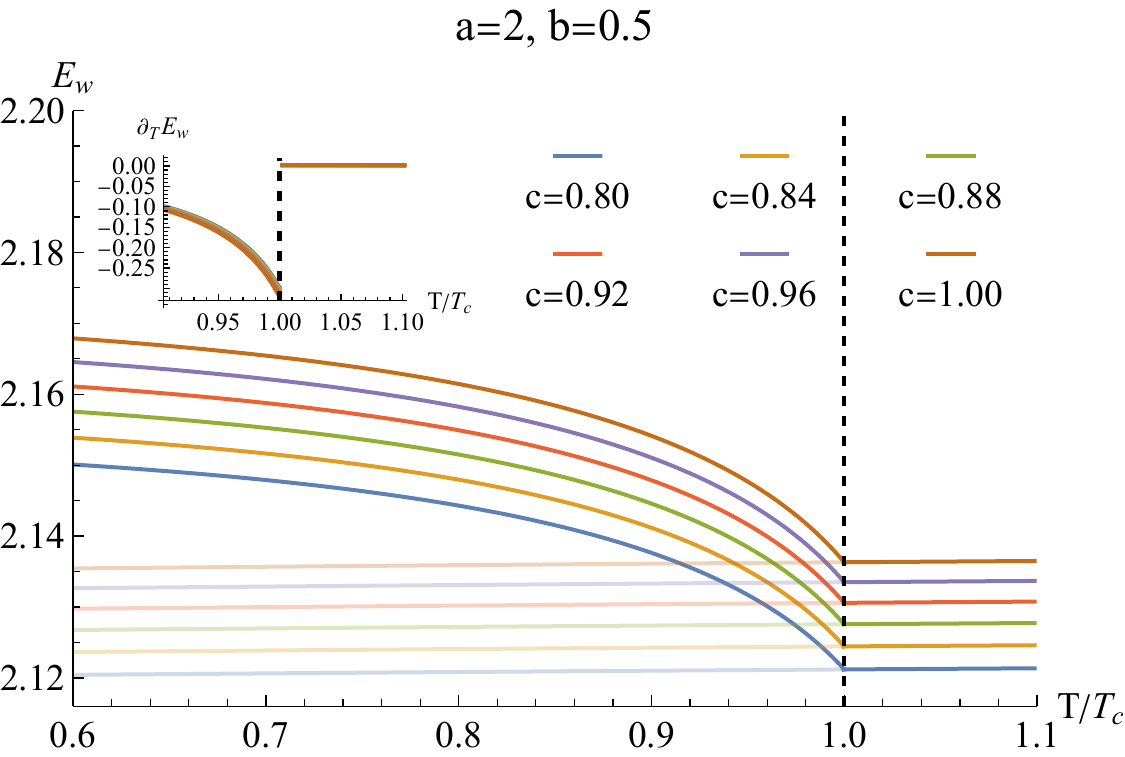}\qquad
  \includegraphics[width=0.45\textwidth]{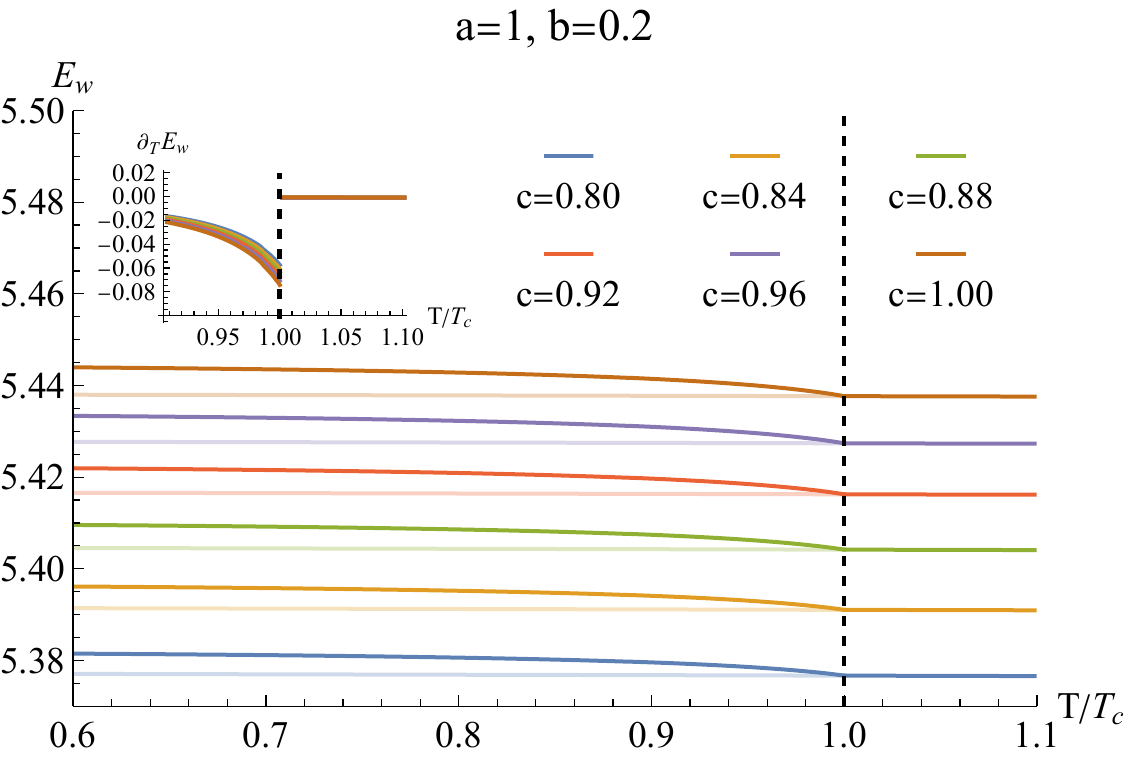}
  \caption{The EWCS $E_w$ vs the temperature $T/T_c$. The inset graph depicts $\partial_TE_w$. The black dashed line depicts the critical temperature. Left plot: The $\partial_TE_w$ of the normal phase is greater than zero when $a=2$ and $b=0.5$. Right plot: The $\partial_TE_w$ of the normal phase is less than zero when $a=1$ and $b=0.2$.}
  \label{fig10}
\end{figure}

We begin by examining the EWCS during a second-order phase transition. Fig. \ref{fig10} shows that EWCS can diagnose the critical behavior of holographic p-wave superconducting phase transitions. At the critical point of a second-order phase transition, EWCS is continuous, but its first derivative is discontinuous. In the superconducting phase, EWCS always decreases with increasing temperature. However, we find that the EWCS in the normal phase is configuration-dependent. In large configurations, it behaves similarly to the HEE, showing a monotonically increasing trend with temperature. In contrast, for small configurations, the EWCS of the normal phase exhibits a monotonically decreasing trend with temperature, opposite to the behavior of the HEE.

\begin{figure}
  \centering
  \includegraphics[width=0.45\textwidth]{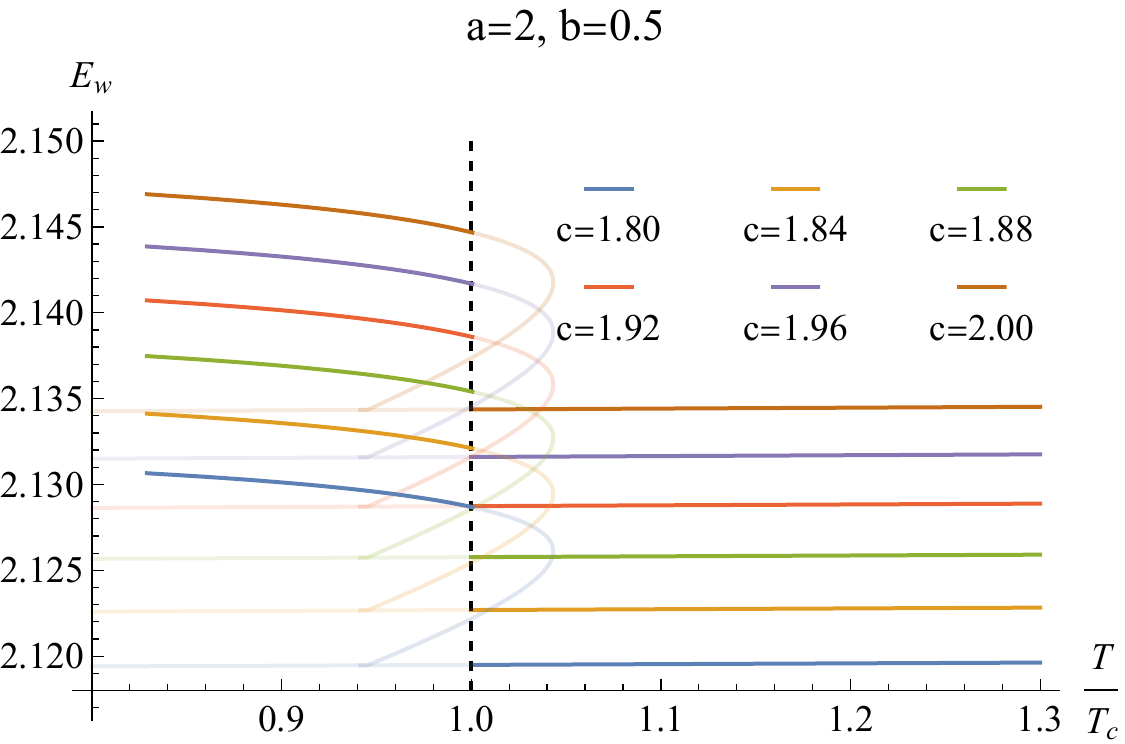}\qquad
  \includegraphics[width=0.45\textwidth]{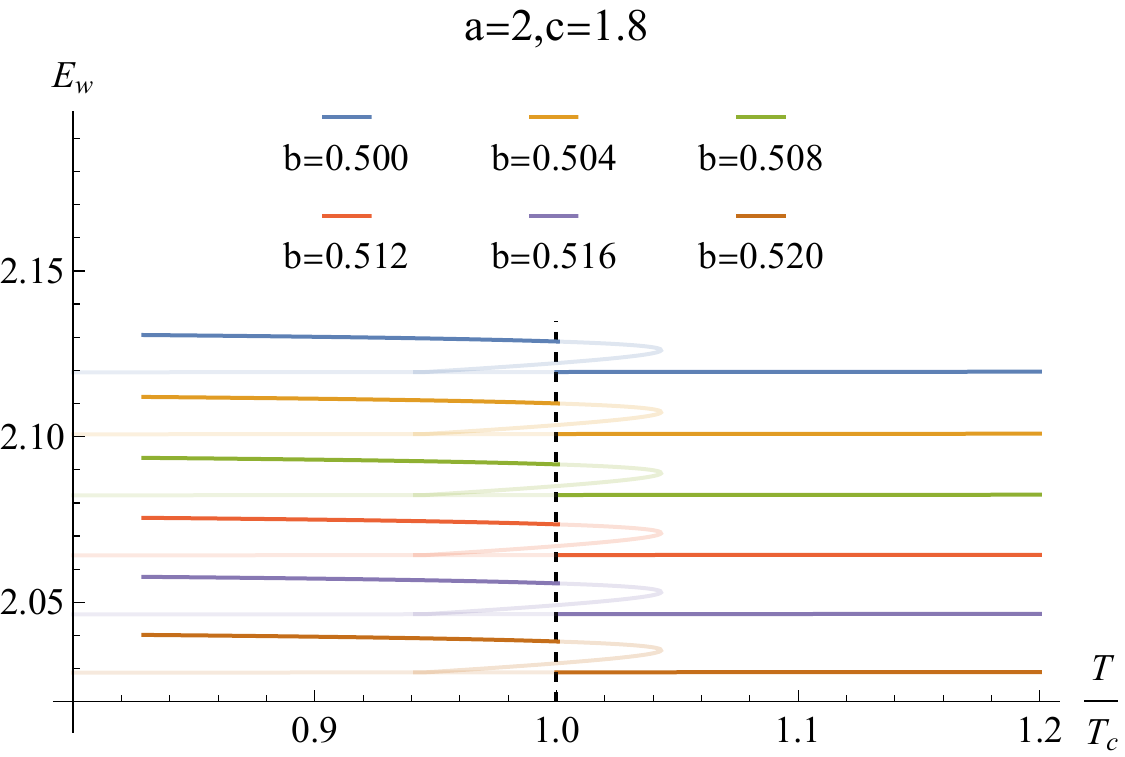}
  \caption{The EWCS $E_w$ versus the temperature $T/T_c$ in the first-order phase transition. The dashed black line represents the critical temperature when $T_c\approx 0.003382$. The translucent line represents the metastable state, whereas the solid line represents the stable state. Left plot: We set $a=2$ and $b=0.5$ with varying $c$ values. Right plot: We set $a=2$ and $c=1.8$ while varying $b$ values.}
  \label{fig11}
\end{figure}

Next, we investigate the behavior of the EWCS during a first-order phase transition. Fig. \ref{fig11} illustrates EWCS behavior during this phase transition, with the inset plot showing the derivative of EWCS with respect to temperature ($\partial_T E_w$) versus temperature $T$. The inset plot illustrates that in normal phase the EWCS decreases with increasing temperature. Unlike the HEE, the EWCS of the superconducting phase always decreases with temperature. When the temperature falls below a critical point, EWCS abruptly jumps from the normal phase to the superconducting phase, this sudden change in EWCS suggests that it can capture the first-order phase transition, similar to the HEE and MI.

In addition to diagnosing the critical points, it is also important to investigate the scaling behavior of the holographic quantum information. Next, we analyze the critical behavior of the quantum information-related quantities during the p-wave superconductivity phase transitions.

\section{The Scaling behavior of the quantum information}\label{sec3}

As the critical point marks the bifurcation point between the normal and superconducting phases, to study the critical behavior, we compare the quantum information quantities of the normal phase to those of the superconducting phase by subtracting the former from the latter,
\begin{equation}
  \delta S_E =S^{\text{cond}}_E-S^{\text{normal}}_E,\quad \delta E_w=E^{\text{cond}}_w-E^{\text{normal}}_w.
\end{equation}
We propose the following critical behaviors for the HEE and EWCS,
\begin{equation}
  \delta S\sim \left(1-\frac{T}{T_c}\right)^{\alpha_{\text{HEE}}},\quad \delta E_w\sim \left(1-\frac{T}{T_c}\right)^{\alpha_{\text{EWCS}}},
\end{equation}
where $\alpha_{\text{HEE}}$ and $\alpha_{\text{EWCS}}$ are the critical exponent of the HEE and EWCS, respectively.
\begin{figure}
  \centering
  \includegraphics[width=0.45\textwidth]{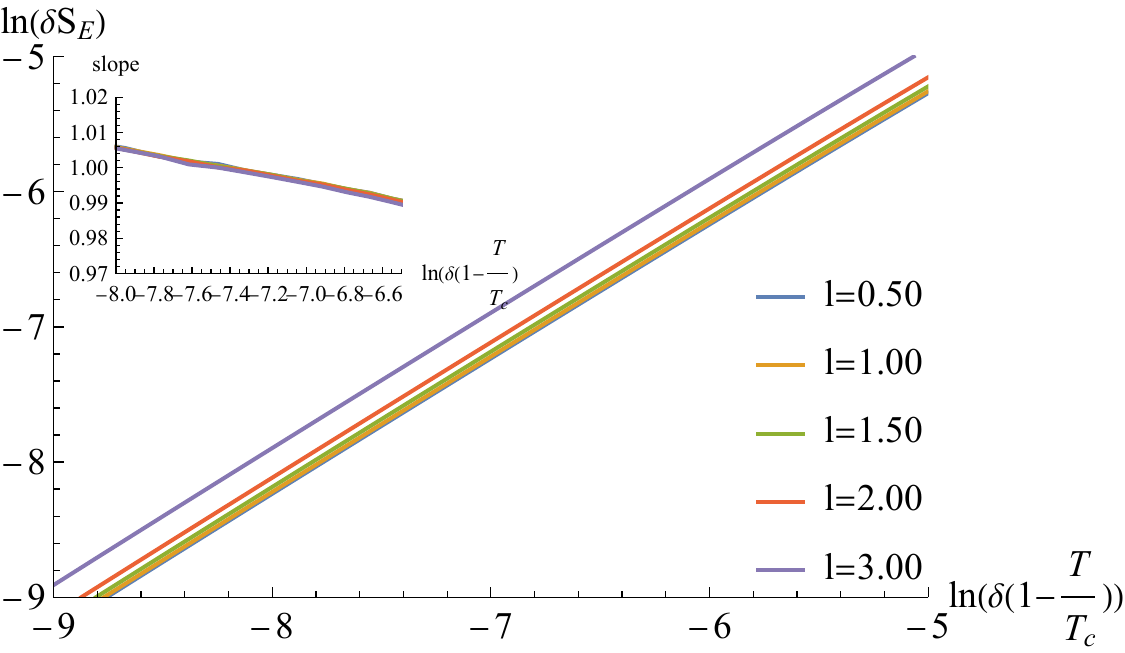}\qquad
  \includegraphics[width=0.45\textwidth]{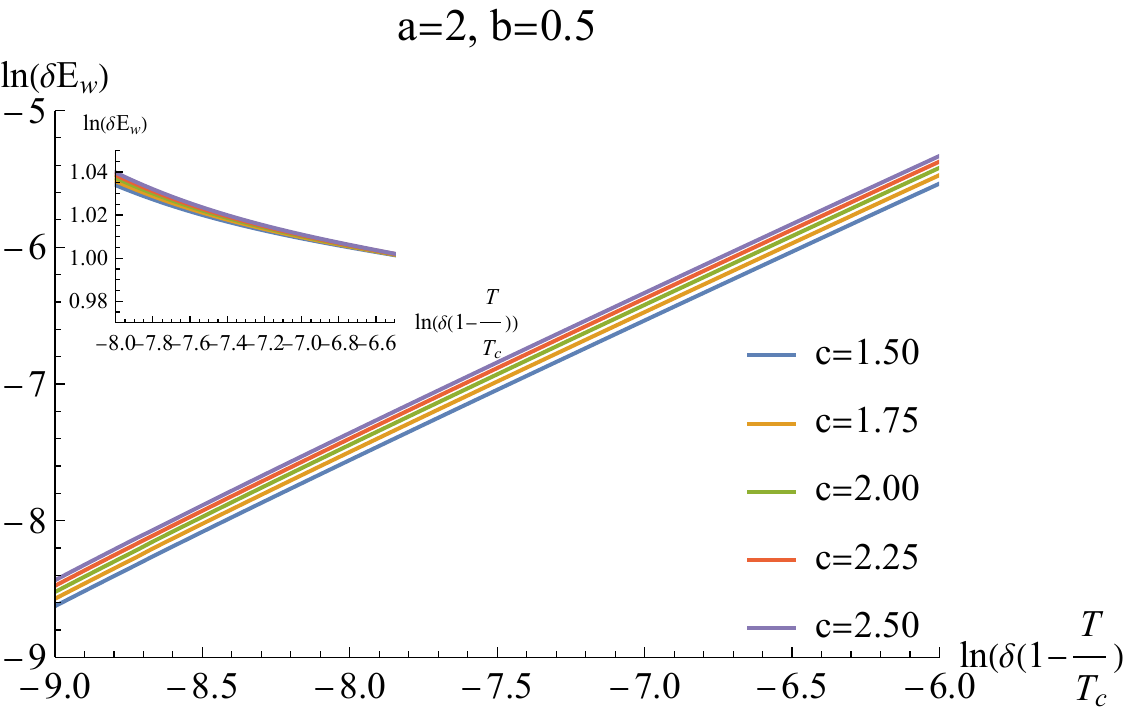}
  \caption{The scaling behavior of HEE and EWCS. The inset plot shows the slope of the holographic quantum information. Left plot: $\text{ln}(\delta S_E)$ versus $\text{ln}(\delta(1-\frac{T}{T c}))$ with different width of $l$. Right plot: we set $a=2$ and $b=0.5$ and $\text{ln}(\delta E_w)$ versus $\text{ln}(\delta(1-\frac{T}{T_c}))$ with different values of $c$.}
  \label{fig12}
\end{figure}
We plot the critical scaling behavior in Fig. \ref{fig12}, from which we find that both EWCS and HEE exhibit excellent scaling behavior near the critical point. More importantly, they both have the same critical exponent,
\begin{equation}
  \alpha_{\text{HEE}}\approx \alpha_{\text{EWCS}}\approx 1.
\end{equation}
It is important to note that the vector field $\rho_\mu$ is always zero at temperatures higher than the critical temperature. At temperatures slightly below the critical point, the condensate vacuum expectation value of $\langle J_x\rangle$ is small and can be analyzed using perturbation theory.
We can expand the vector field $\rho_\mu$ and the metric function near the critical point as \cite{Zeng:2010zn,Pan:2012jf,Ammon:2009xh},
\begin{equation}\label{eq:expansion}
  \begin{aligned}
    \rho_{x} & =\epsilon\rho^{(1)}+\epsilon^3\rho^{(3)}+\epsilon^5\rho^{(5)}+\cdots, \\
    U        & =1+\epsilon^2 U^{(1)}+\epsilon^4 U^{(4)}+\cdots,                      \\
    V_1      & =1+\epsilon^2 V_1^{(1)}+\epsilon^4 V_1^{(4)}+\cdots.
  \end{aligned}
\end{equation}
From \eqref{eq:expansion}, we can deduce that the critical exponent of the metric function $U$, $V_1$, is twice that of the condensate $\langle J_x\rangle$. This can be understood by noting that holographic quantum information is represented by geometric objects that depend only on the metric. Their critical exponent can be written as,
\begin{equation}
  \delta(S_E)\sim\delta(E_w)\sim\delta(\langle J_x\rangle)^2\sim\left(1-\frac{T}{T_c}\right)^{2\alpha_c}.
\end{equation}
Therefore, the theoretical critical exponent of holographic quantum information should be twice that of the condensate $\langle J_x\rangle$,
\begin{equation}
  \alpha_{\text{HEE}}=\alpha_{\text{EWCS}}= 2\alpha_c.
\end{equation}

Although EWCS and HEE have the same critical exponent in the critical region, they do not tend to the scaling law at the same rate in the critical region. To better investigate this phenomenon near the critical point, we define the quasi-critical exponent (QCE) as
\begin{equation}
  \alpha \equiv \frac{d \ln (\delta S)}{d \ln \left(1-\frac{T}{T_c}\right)}.
\end{equation}
QCE is a function of $\ln\left(1-\frac{T}{T_c}\right)$. Apparently, the QCE $\alpha$ behavior along $\ln\left(1-\frac{T}{T_c}\right)$ can measure the extent to which a wide range of $S$ can converge to the scaling law.

\begin{figure}
  \centering
  \includegraphics[width=0.45\textwidth]{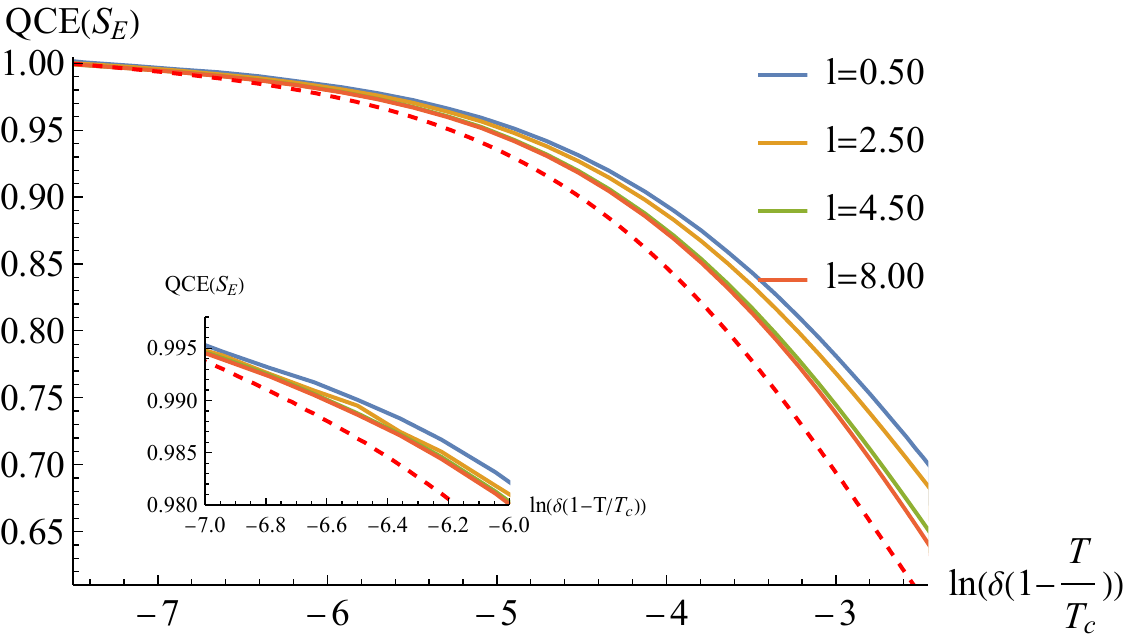}\qquad
  \includegraphics[width=0.45\textwidth]{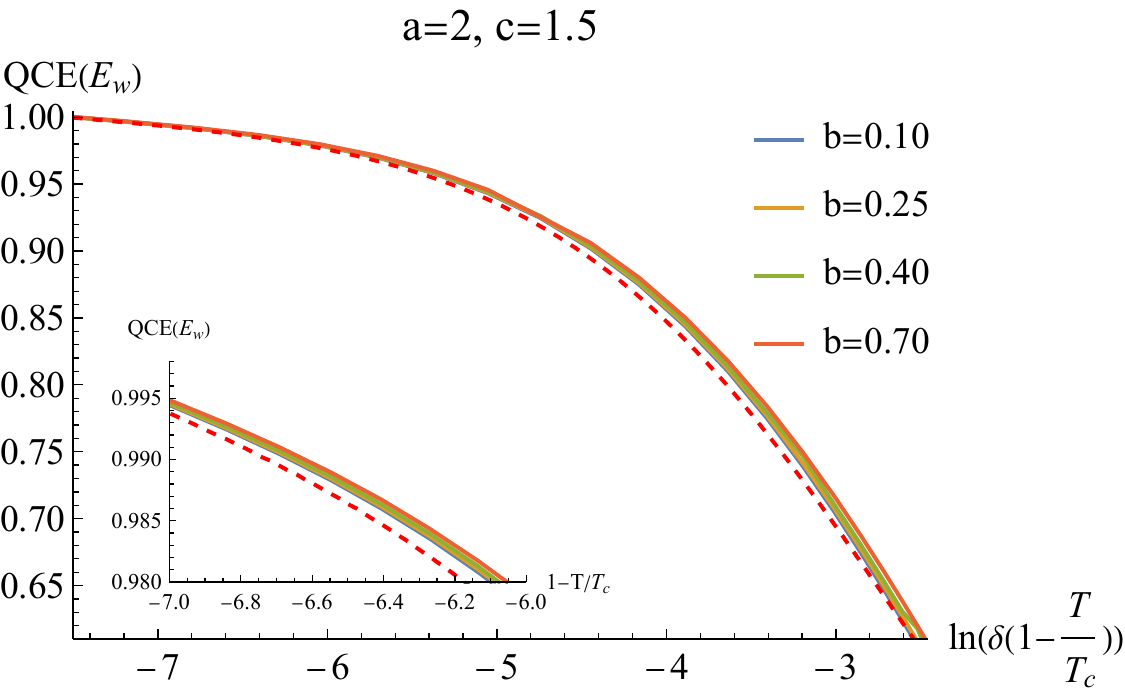}
  \caption{The QCE of the HEE and EWCS. The red dashed line represents the twice QCE of the condensate $\langle J_x\rangle$. Left plot: The QCE of the HEE near the critical points. Right plot: The QCE of the EWCS near the critical points, when we fix the $a=2$ and $c=1.5$.}
  \label{fig13}
\end{figure}

We show the QCE of HEE and EWCS in Fig. \ref{fig13}.
From the left plot of Fig. \ref{fig13} we find that the width $l$ has an impact on the scaling behavior of HEE. As the width $l$ increases, the scaling behavior of HEE is closer to the theoretical scaling behavior. As the temperature moves away from the critical point or the width $l$ decreases, however, the scaling behavior of the HEE begins to deviate from the theoretical result.

The QCE of EWCS is depicted in the right plot of Fig. \ref{fig13}. Comparing the left plot and the right plot of Fig. \ref{fig13}, EWCS converges to the theoretical scaling law over a broader range. As the separation $b$ decreases, the scaling behavior of EWCS becomes close to the theoretical results. This behavior suggests that EWCS, as a measure for mixed-state entanglement, can more accurately describe the scaling behavior during superconductivity phase transitions than HEE.

\section{The growth rate of the holographic quantum information}\label{sec4}

Several important inequalities involving the EWCS have been proposed in the literature \cite{Umemoto:2018jpc,Li:2022wim,Bao:2017nhh}, such as the inequality $E_w(\rho_{AC})\geq \frac{1}{2}I(A,c)$, which states that the EWCS cannot be smaller than half of the MI. These inequalities are crucial in the study of mixed-state entanglement measures, particularly in testing the validity of holographic duals of certain quantum information. In this paper, we find a new inequality behavior of EWCS and MI related to the superconductivity phase transition: near the phase transition point, the relative growth rate of MI along the temperature axis is always greater than that of EWCS.

\begin{figure}
  \centering
  \includegraphics[width=0.45\textwidth]{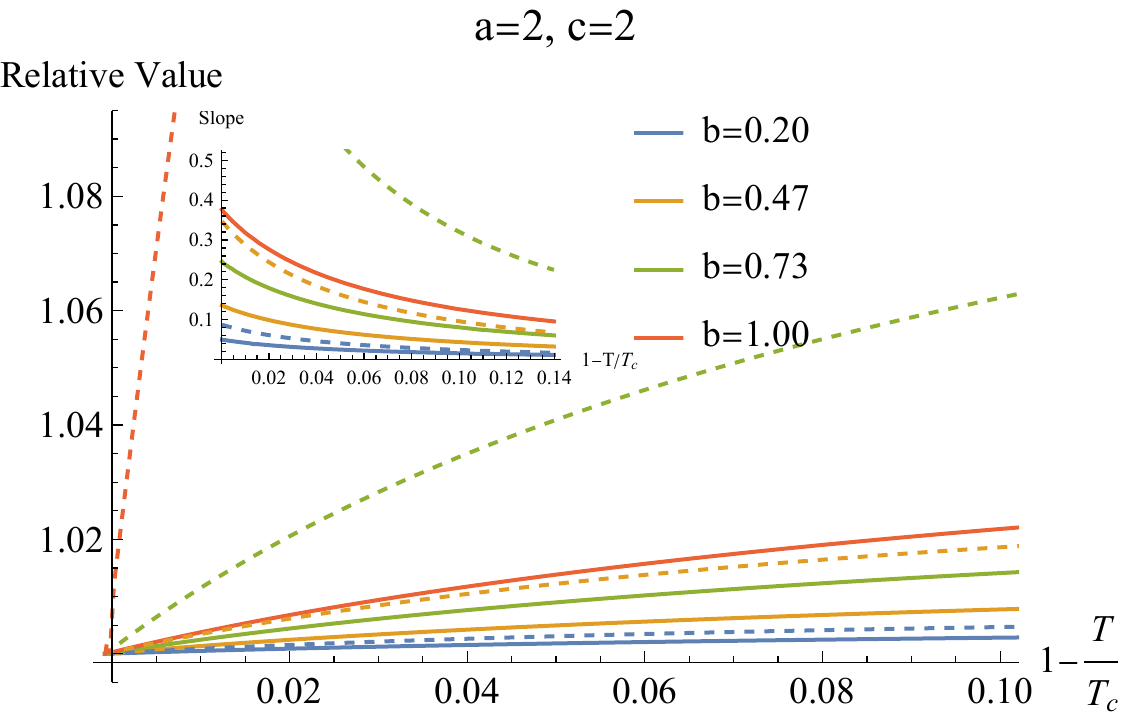}\qquad
  \includegraphics[width=0.45\textwidth]{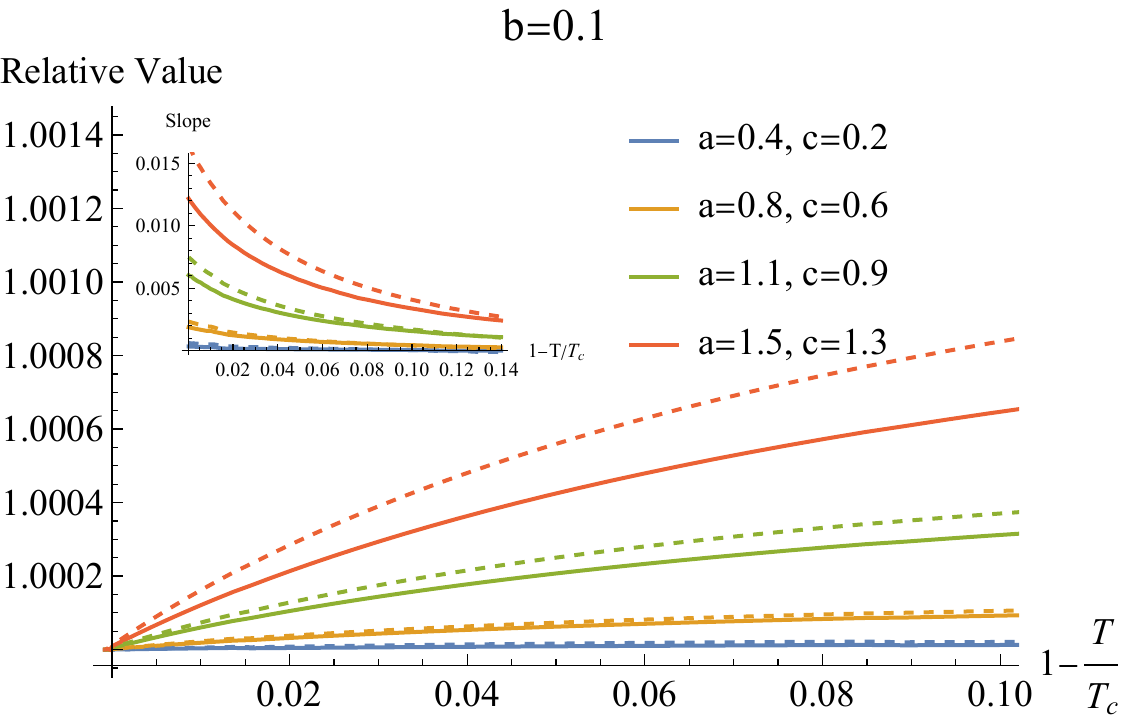}
  \caption{The relative value of MI and EWCS near the critical point. The dashed lines represent $\tilde{I}$ and the solid lines represent $\tilde{E}_w$. The inset plot is the slope of $\tilde{I}$ and $\tilde{E}_w$. Left plot: We fix the subsystems $a=c=2$ and change the separation $b$. Right plot: We fix the separation $b=0.1$ and change the subsystem $a$ and $c$.}
  \label{fig15}
\end{figure}

When the temperature drops below the critical temperature, the EWCS and the MI of the superconducting phases are always larger than those of the normal phases. To take a closer look at the relationship between the EWCS and the MI, we define the relative values of the MI and the EWCS,
\begin{equation}
  \tilde{E}_w=\frac{E_{w,\text{cond}}}{E_{w,\text{norm}}},\quad \tilde{I}=\frac{I_{\text{cond}}}{I_{\text{norm}}}.
\end{equation}
With this definition, $\tilde{E}_w$ and $\tilde{I}$ are fixed at $1$ at the critical point. In Fig. \ref{fig15}, we depict the relationship between the $\tilde{I}$ and $\tilde{E}_w$. Contrary to the inequality \cite{Umemoto:2018jpc,Li:2022wim,Bao:2017nhh}, the relative MI is always larger than the relative values of EWCS in the critical region. To describe this relationship quantitatively, we examine the fact that,
\begin{equation}\label{eq:dqdt}
  \delta(Q) \simeq A(Q)\left(1-\frac{T}{T_c}\right)^{\alpha},
\end{equation}
where $Q$ stands for any physical quantity possessing critical behaviors. From \eqref{eq:dqdt} we find that,
\begin{equation}\label{eq:qot}
  \tilde E_w = 1 + A(\tilde E_w) \left(1-\frac{T}{T_c}\right)^{\alpha}, \quad \tilde I = 1 + A(\tilde I) \left(1-\frac{T}{T_c}\right)^{\alpha}.
\end{equation}
Accordingly, it can be seen that $ A $ actually measures the increasing phenomenon of holographic quantum information in Fig. \ref{fig13}, and hence we call $ A $ the growth rate. We work out $A(\tilde E_w)$ and $A(\tilde I)$ for several different configurations and list them in Table \ref{table:aa}. From these numerical results we conclude a new inequality between the EWCS and MI growth rates near the critical point,
\begin{equation}\label{eq:ineq2}
  A(\tilde{I})>A(\tilde{E}_w).
\end{equation}
The growth rate of MI is always greater than that of EWCS near the critical point. Furthermore, the difference between the growth rates of EWCS and MI increases as the subsystem separation $b$ increases. Near the critical point, the entanglement of the system changes rapidly, and MI is more sensitive to these changes than EWCS. This tendency could be attributed to MI's ability to capture the total correlation of the system, which exceeds the information captured by EWCS. Additionally, we have examined this inequality in other models of thermal phase transitions, including the holographic s-wave superconductor model, and propose that this inequality may be universal in thermal phase transitions.

\begin{table}
  \caption{The growth rate $A(\tilde{E}_w)$ and $A(\tilde{I})$ at different configurations.}
  \label{table:aa}
  \begin{tabular}{|m{6cm}<{\centering}|m{2cm}<{\centering}|m{2cm}<{\centering}|}
    \hline
    Configuration  \qquad              & A($\tilde{E}_w$) & A($\tilde{I}$) \\ \hline\hline
    $a=c=2$, $b=0.20$  \qquad          & 0.0426           & 0.0741         \\ \hline
    $a=c=2$, $b=0.47$  \qquad          & 0.1162           & 0.2947         \\ \hline
    $a=c=2$, $b=0.73$  \qquad          & 0.2094           & 0.9855         \\ \hline
    $a=c=2$, $b=1.00$  \qquad          & 0.3212           & 10.8712        \\ \hline
    $a=0.8$, $b=0.2$, $c=0.4$  \qquad  & 0.00218          & 0.00525        \\ \hline
    $a=0.8$, $b=0.2$, $c=0.8$   \qquad & 0.00520          & 0.00803        \\ \hline
    $a=0.8$, $b=0.2$, $c=1.2$   \qquad & 0.00862          & 0.01336        \\ \hline
    $a=0.8$, $b=0.2$, $c=1.6$   \qquad & 0.01209          & 0.01975        \\ \hline
    $a=0.5$, $b=0.2$, $c=0.4$  \qquad  & 0.00116          & 0.00364        \\ \hline
    $a=1.0$, $b=0.2$, $c=0.9$   \qquad & 0.00794          & 0.01182        \\ \hline
    $a=1.5$, $b=0.2$, $c=1.4$   \qquad & 0.02016          & 0.03120        \\ \hline
    $a=3.0$, $b=0.2$, $c=2.9$   \qquad & 0.06042          & 0.11959        \\ \hline
  \end{tabular}
\end{table}

\section{Discussion}\label{sec5}

In this study, we investigate mixed-state entanglement measures, including HEE, MI and EWCS, in a holographic p-wave superconductor model. The model exhibits both second and first-order phase transitions when varying system parameters. We find that HEE and EWCS can accurately diagnose the critical behavior of these phase transitions. Additionally, we observe that the behavior of HEE is related to thermodynamic entropy as the subsystem configuration increases. However, as a mixed-state entanglement measure, EWCS exhibits the opposite behavior from HEE in the superconducting phase. Specifically, HEE always increases with temperature, whereas EWCS in the superconducting state decreases with temperature. In the case of first-order phase transitions, the holographic quantum information experiences sudden changes. However, the EWCS behavior in the normal phase is dependent on the subsystem configuration. This behavior demonstrates that EWCS can not only detect phase transitions but also capture more information than HEE.

In addition to diagnosing phase transitions, we also examine the scaling behaviors of the condensate and the holographic quantum information. Through analyzing the scaling behavior of various holographic quantum information measures, we find that HEE and EWCS not only detect the critical point but also exhibit scaling behaviors. We show both numerically and analytically that the critical exponent of holographic quantum information is twice that of the condensate. Furthermore, we observe that compared to HEE, EWCS provides a more sensitive characterization of the scaling behavior, making it more suitable as a measure for mixed-state entanglement in superconductivity phase transitions. Additionally, we propose a novel inequality for EWCS and MI in phase transitions and provide numerical evidence for this result. The relative growth rate of MI is always larger than that of EWCS near the critical point.

Next, we point out several directions worth further investigation.
The investigation of topological and quantum phase transitions is an important area of research in condensed matter theory \cite{Donos:2013eha,Landsteiner:2015pdh,Ling:2016dck,Baggioli:2018afg,Baggioli:2020cld}. In addition, the relationship between HEE and quantum phase transitions has been studied under holographic framework in previous works \cite{Ling:2016wyr,Ling:2015dma}. Further research into the mixed-state entanglement in quantum phase transitions and topological quantum phase transitions is therefore desirable. Additionally, it would be interesting to test the inequality \eqref{eq:ineq2} in other thermal phase transition models, such as the $d$-wave superconductivity model and the massive gravity model. We are working on these directions.

\section*{Acknowledgments}\label{sec:ack}

Peng Liu would like to thank Yun-Ha Zha for her kind encouragement during this work. Zhe Yang appreciates Feng-Ying Deng's support and warm words of encouragement during this work. We are also very grateful to Chong-Ye Chen, Mu-Jing Li, and Wei Xiong for their helpful discussion and suggestions. This work is supported by the Natural Science Foundation of China under Grant No. 11905083, 12005077 and 11805083, as well as the Science and Technology Planning Project of Guangzhou (202201010655) and Guangdong Basic and Applied Basic Research Foundation (2021A1515012374).

\end{document}